\documentclass[letters,usenatbib]{mnras}

\usepackage[T1]{fontenc}

\DeclareRobustCommand{\VAN}[3]{#2}
\let\VANthebibliography\thebibliography
\def\thebibliography{\DeclareRobustCommand{\VAN}[3]{##3}\VANthebibliography}

\usepackage{graphicx}	
\usepackage{amsmath}	
\usepackage{amssymb}	
\usepackage{newtxtext,newtxmath}
\usepackage{soul,color}

\title[Solar wind helium abundance]{Evidence for distinctive changes in the solar wind helium abundance in cycle 24 }

\author[Yogesh et al.]{
Yogesh,$^{1,2}$\thanks{E-mail: yphy22@gmail.com}
D. Chakrabarty$^1$,
and N. Srivastava$^3$
\\
$^{1}$Physical Research Laboratory, Navrangpura, Ahmedabad 380009, India\\
$^{2}$Indian Institute of Technology-Gandhinagar, Gandhinagar 382424, India\\
$^{3}$Udaipur Solar Observatory, Physical Research Laboratory, Udaipur -313001, India
}

\date{Accepted 2021 February 08. Received 2021 February 02; in original form 2020 January 04}

\pubyear{2020}

\begin{document}
\label{firstpage}
\pagerange{\pageref{firstpage}--\pageref{lastpage}}
\maketitle

\begin{abstract}
The relative abundance of alpha particles with respect to proton, usually expressed as  $A_{He}$ = ($n_\alpha/n_p$)*100, is known to respond to solar activity although changes in its behaviour in the last four solar cycles are not known. In this letter, by systematically analysing inter-calibrated $A_{He}$ data obtained from the first Lagrangian point of the Sun-Earth system, we show that $A_{He}$ variations are distinctively different in solar cycle 24 as compared to the last three cycles. The frequency of $A_{He}$ = 2-3\% events is significantly higher in slow/intermediate solar winds in cycle 24 as opposed to the dominance of the typical $A_{He}$ = 4-5\% events in the previous three cycles. Further, the occurrence of $A_{He}$ $\geq$ 10\% events is significantly reduced in cycle 24. Not only that, the changes in delay of $A_{He}$ with respect to peak sunspot numbers are less sensitive to changes in solar wind velocity in cycle 24. The investigation suggests that the coronal magnetic field configuration started undergoing systematic changes starting from cycle 23 and this altered magnetic field configuration affected the way helium got processed and depleted in the solar atmosphere.

\end{abstract}

\begin{keywords}
Sun: abundances -- Sun: activity -- solar wind
\end{keywords}



\section{Introduction}
Doubly ionized helium ($He^{2+}$ or $\alpha$ particle) is the second most abundant species in the solar wind after the singly ionized Hydrogen ($H^+$ or proton). Despite helium being the second most abundant element in the solar wind, doubly ionized helium atom abundance in the quiet solar wind does not exceed 4-5\% \citep{Laming2001Feldman} and is strongly regulated by a number of processes occurring in the chromosphere, transition region and corona. Helium is four times heavier than hydrogen and as a consequence, it constitutes 25\% of the solar wind mass flux. The first ionization potential (FIP) of helium being the highest among the solar elements, helium gets ionized later than any other element and this occurs at the topside of the chromosphere. It is believed that the FIP effect \citep{Laming2015} primarily depletes the helium abundance in the chromosphere and the transition region. Processes like gravitational settling, Coulomb collisions and other wave-particle interactions also contribute in varying degrees to the helium depletion process at different heights \citep{Moses2020}.

The variation of $A_{He}$ with solar activity was first indicated by \cite{Hirshberg1973} and \cite{Ogilvie1974} by comparing these with the variations in the sunspot numbers (SSN). These authors used multiple satellite data to show the relationship between $A_{He}$ and SSN. Subsequently, \cite{feldman1978long} used IMP satellite data to extend the result of \cite{Ogilvie1974} and indicated a possible delay between the variations in $A_{He}$ and SSN. \cite{Aellig2001} extended the solar cycle variation of $A_{He}$ using WIND data and found out a linear relationship between the SSN and the $A_{He}$ for the slow solar wind. No such strong relationship was found to exist for the fast-solar wind. It was also found that the linear relationship between the solar wind velocity and the $A_{He}$ for the slow solar wind is strong during minima but weak during maxima. Further, in order to understand the variation of $A_{He}$ for slow solar wind, \cite{Kasper2007, Kasper2012} investigated the solar activity variation of $A_{He}$ for approximately one full solar cycle. The above studies suggest the possible association between the sources of the solar wind and solar activity variation of $A_{He}$. In addition to this, right from the work of \cite{feldman1978long}, it is suggested that there exists a delay (hysteresis) between $A_{He}$ and SSN. In recent times, \cite{Alterman2019} investigated the delay in $A_{He}$ variation with respect to the variations in SSN and found a linear relationship between the solar wind velocity and the delay. Although the solar activity dependence of $A_{He}$ and the hysteresis between $A_{He}$ and SSN were shown, it is not known whether helium abundance in solar wind behaved identically in the last four solar cycles. The answer to this question can shed light on the long-term changes in the way helium got processed in the coronal magnetic field of the Sun. This is an important question as many authors showed declining solar activity from cycle 23 onwards \citep{Janardhan2011}. The results presented in this work provide evidence that the variation of helium abundance started changing from cycle 23 onwards and is conspicuously different in cycle 24.  

\begin{figure*}
\includegraphics[scale=0.32]{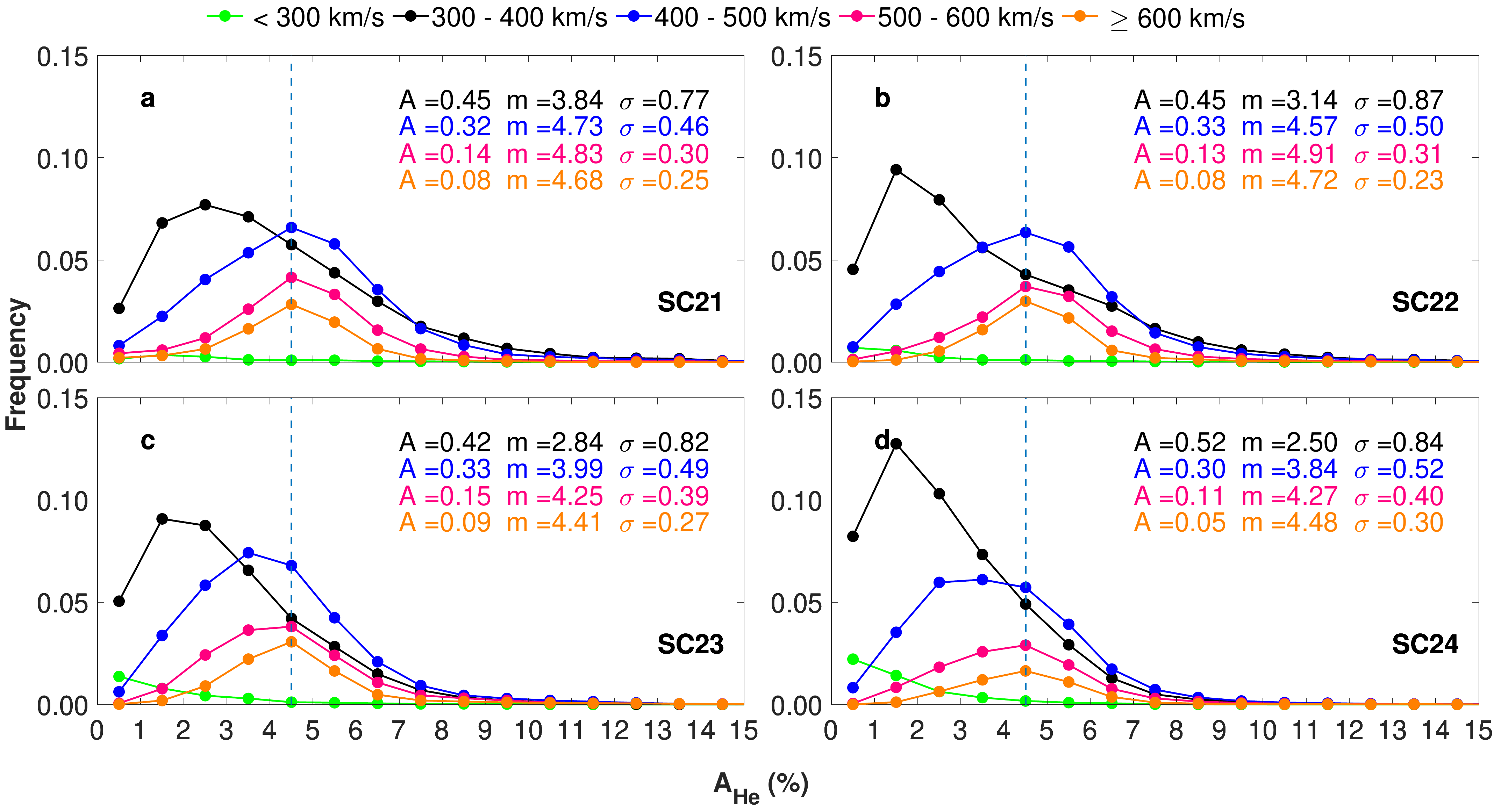}
\caption{\textbf{Frequency of distribution of $A_{He}$ in different velocity bins in the last four solar cycles.}  Frequency distribution of $A_{He}$ events with 1\% bin size for four velocity bins, viz. (1) $<$ 300 km/s (2) 300-400 km/s  (3) 400-500 km/s (4) 500-600 km/s (6) $\geq$ 600 km/s in solar cycles 21 (a), 22 (b), 23 (c) and 24 (d). Each velocity window is marked with different colors (filled circle joined by solid line) and are shown at the top of the Figure. The frequency distributions are approximated by log-normal distributions wherein A, $\sigma$ and m are the fit parameters namely normalization coefficient, standard deviation and median in linear scale respectively. It can be noted that m monotonically decreases for 300-400 km/s and 400-500 km/s velocity bin as one goes from cycle 21 to cycle 24. Further, the shift of the peak $A_{He}$ from the usual 4-5\% (in cycles 21 and 22) to 2-3\% in the 400-500 km/s velocity range and increase in the frequency in the 300-400 km/s velocity range in cycle 24 are also conspicuous. The vertical dashed lines mark $A_{He}$ of 4.5\%. This figure also reveals that the frequency of occurrence for $A_{He}$ $>$ 10\% events are insignificant in statistical sense. \label{fig:1}}
\end{figure*}

\section{Dataset}
The low resolution OMNI (LRO) database used in this work contains multi-spacecraft data (like IMP, Wind, ACE, Geotail etc.,) having 1-hour cadence. Some of these satellites are in earth orbit and for these satellites, the measurements made outside the terrestrial magnetosphere are considered. The plasma data from some spacecraft and parameters are compared and cross-normalized with respect to the Wind/SWE/NLF data. For the present investigation wherein proton and alpha measurements are used, primarily four multi-source data are used. These are IMP6/IMP7/IMP8 satellite data during 1971-1978, IMP8 data during 1973-2001, IMP8/ISEE3 data during 1978-1982 and IMP8, Wind, ACE, Geotail data during 1995-2019. These aspects are detailed comprehensively in \url{https://omniweb.gsfc.nasa.gov/html/ow_data.html}. One can also find the discussions on the time shifting of the dataset for various satellites, data averaging scheme, spacecraft prioritization, cleaning of source data as well as the possible sources of random and systematic differences between hourly averages of pairs of like parameters obtained by two spacecraft in this link. In addition, \cite{King2005} and references cited therein also provide the nuances of the normalization scheme for ACE and WIND satellites that are major contributors to this ONMI dataset. Although the OMNI database starts from 1963, the $A_{He}$ dataset is available since 1971. The data are parsed to create data for the solar cycles 21-24 marked by the years 1976-1986, 1987-1996, 1997-2008 and 2009-2019 respectively.

\section{Results}
Solar cycle 24 turns out to be the weakest cycle in the last one hundred years \citep{hathaway2015solar}. In order to evaluate the changes, if any, in the solar wind helium  abundance in cycle 24 compared to the previous three cycles, we performed a number of extensive analyses of $A_{He}$ data taken at  1-hour cadence and available in the OMNI database (\url{https://cdaweb.gsfc.nasa.gov/index.html/}) spanning over almost half a century and  encompassing the last four solar cycles (cycle 21 to cycle 24).
\subsection{Frequency of $A_{He}$ events ($\leq$ 10\% and $\geq$ 10\%) in the last four solar cycles  }

As a first step, we divided the 1 hourly $A_{He}$ data for each solar cycle in five velocity bins, viz. (1) $<$ 300 km/s, (2) 300-400 km/s, (3) 400-500 km/s, (4) 500-600 km/s, and (5) $\geq$ 600 km/s. For a given velocity bin, we evaluated the frequency distributions of $A_{He}$ events. We note that the frequency distributions can be well-approximated by log-normal distributions for all the velocity bins except for the bin $<$ 300 km/s (Figure:\ref{fig:1}).
\begin{equation}
Frequency = \frac{A}{\sigma \sqrt{2 \pi}} exp \{ \frac{-(log(A_{He})-\mu)^2}{2 \sigma^2}\}
\label{eq1}
\end{equation}
\begin{equation}
m = exp (\mu)
\label{eq2}
\end{equation}
                             
In equation (\ref{eq1}), A is the normalization factor, $\sigma$ is the standard deviation (shape parameter) and $\mu$ is the centre (median) of the log-normal distribution. Equation (\ref{eq2}) is used to convert $\mu$ from log scale to linear scale. This linear counterpart of $\mu$ is marked as m and this is basically the median value. The corresponding values of A, m and $\mu$ are mentioned in each sub-plot of Fig. \ref{fig:1}. It can be noted that the median value of $A_{He}$ monotonically decreases for 300-400 km/s and 400-500 km/s velocity bin as one goes from cycle 21 to cycle 24. However, for the other two velocity bins, the median value increases in cycle 22 before dropping again in cycles 23 and 24. The most important feature to be noted is the shift of the peak $A_{He}$ from the usual 4-5\% (in cycles 21 and 22) to 2-3\% in the 400-500 km/s velocity range and increase in the frequency in the 300-400 km/s velocity range in cycle 24. Interestingly, these two velocity bins contain around $\sim$ 80\% of the total data points. Based on Fig. \ref{fig:1}, we fix the upper limit of $A_{He}$ at 10\% and proceed to evaluate the velocity dependence of $A_{He}$ for the four solar cycles in sections 3.2 and 3.3.
 \begin{figure*}
\includegraphics[scale=0.32]{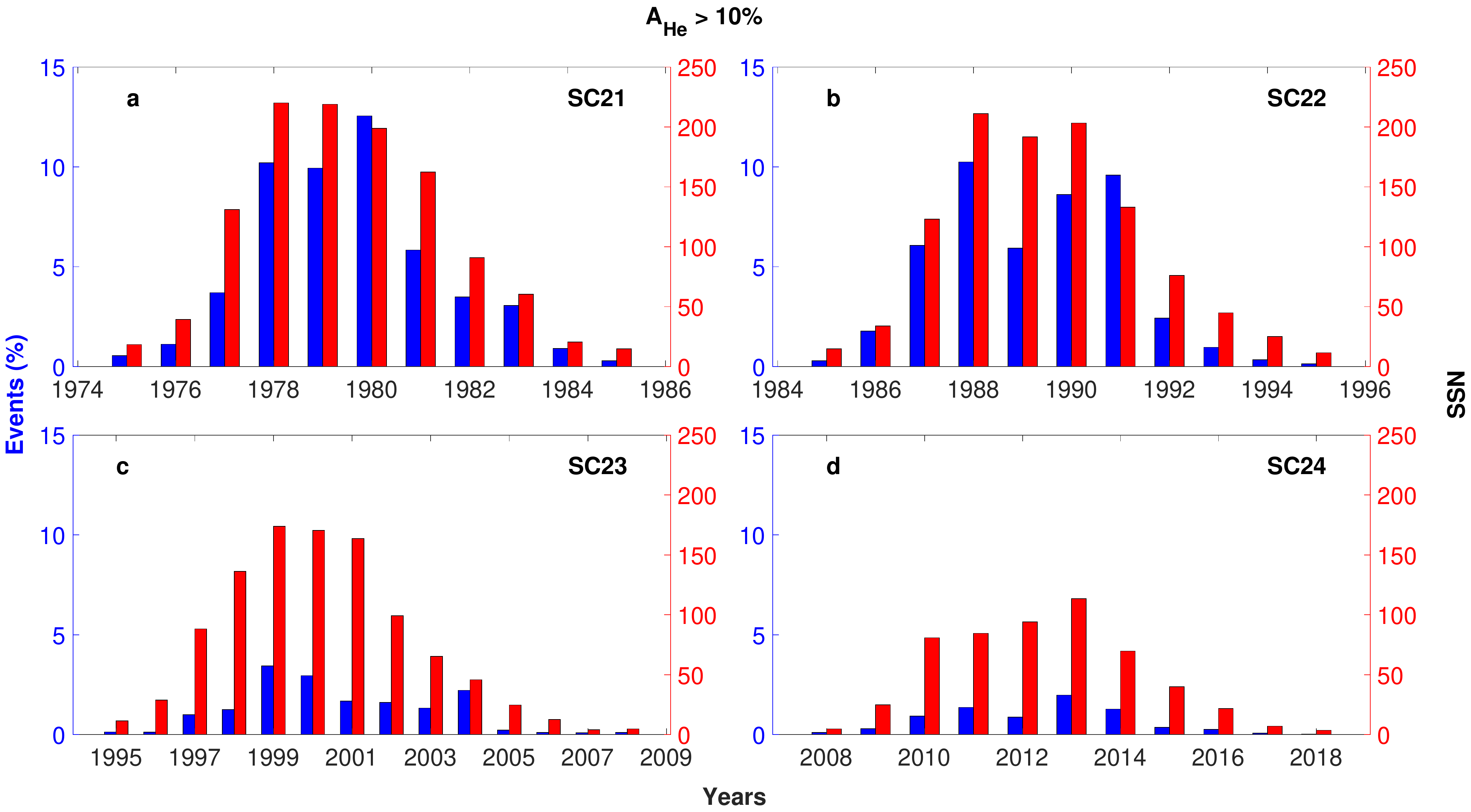}
\caption{\textbf{Frequency of $A_{He}$ $>$ 10\% events in the last four cycles.} Year-wise frequency of $A_{He}$ $>$ 10\% events (marked in blue vertical bars) and yearly averaged sunspot numbers (SSN) (marked in red vertical bars) in the last four solar cycles marked by a, b, c and d. This frequency is calculated with respect to the number of all $A_{He}$ events in a particular year. Peak value of the $A_{He}$ frequencies are 12.55\%, 10.25\%, 3.44\% and 1.98\% as one goes from cycle 21 to 24. \label{fig:2}}
\end{figure*}

We also evaluate the yearly frequency of events with $A_{He}$ $>$ 10\% vis-à-vis yearly averaged sunspot numbers (SSN) for the four solar cycles (Fig.\ref{fig:2}) and note that these events start reducing significantly from cycle 23 and by cycle 24, the frequency of these events does not exceed 2\%. 

\subsection{Solar cycle variation of $A_{He}$ $\leq$ 10\%}

 \begin{figure*}
\includegraphics[scale=0.32]{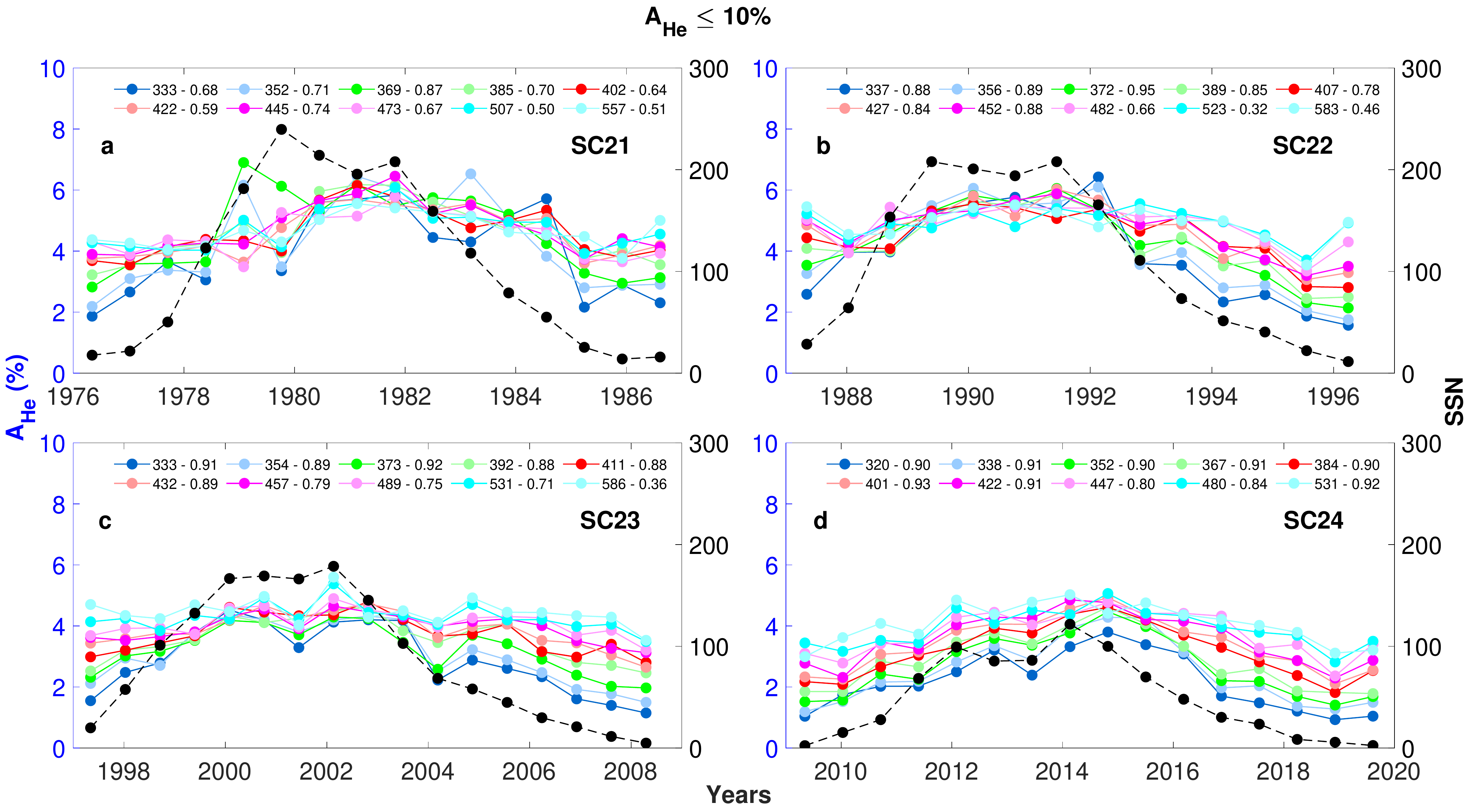}
\caption{\textbf{Variation of $A_{He}$ $\leq$ 10\% in the last four solar cycles: -} Spearman Rank Correlation analysis between averaged (for 250 days) $A_{He}$ and sunspot numbers (SSN) for 12 velocity quantiles in the last four solar cycles. Results for cycle 21 to cycle 24 are represented in a, b, c and d respectively. The Spearman Rank Correlation coefficient, $\rho$ ($A_{He},SSN$), values and the mid-point of each velocity quantile are shown on the top of each panel. Note that the highest and lowest velocity quantiles are removed from this representation. Each velocity quantiles are marked by filled circles of different colours joined by solid lines and SSN variations are marked by black filled circles joined by dashed lines. The $\rho$ ($A_{He},SSN$) values maximize at 369 km/s, 372 km/s, 373 km/s in cycles 21, 22 and 23 respectively. However, in cycle 24, The $\rho$ ($A_{He},SSN$) value maximizes at 401 km/s. The apparent phase offset between the $A_{He}$ and SSN becomes negligible in cycle 24. \label{fig:3}}
\end{figure*}

 \begin{figure*}
\includegraphics[scale=0.32]{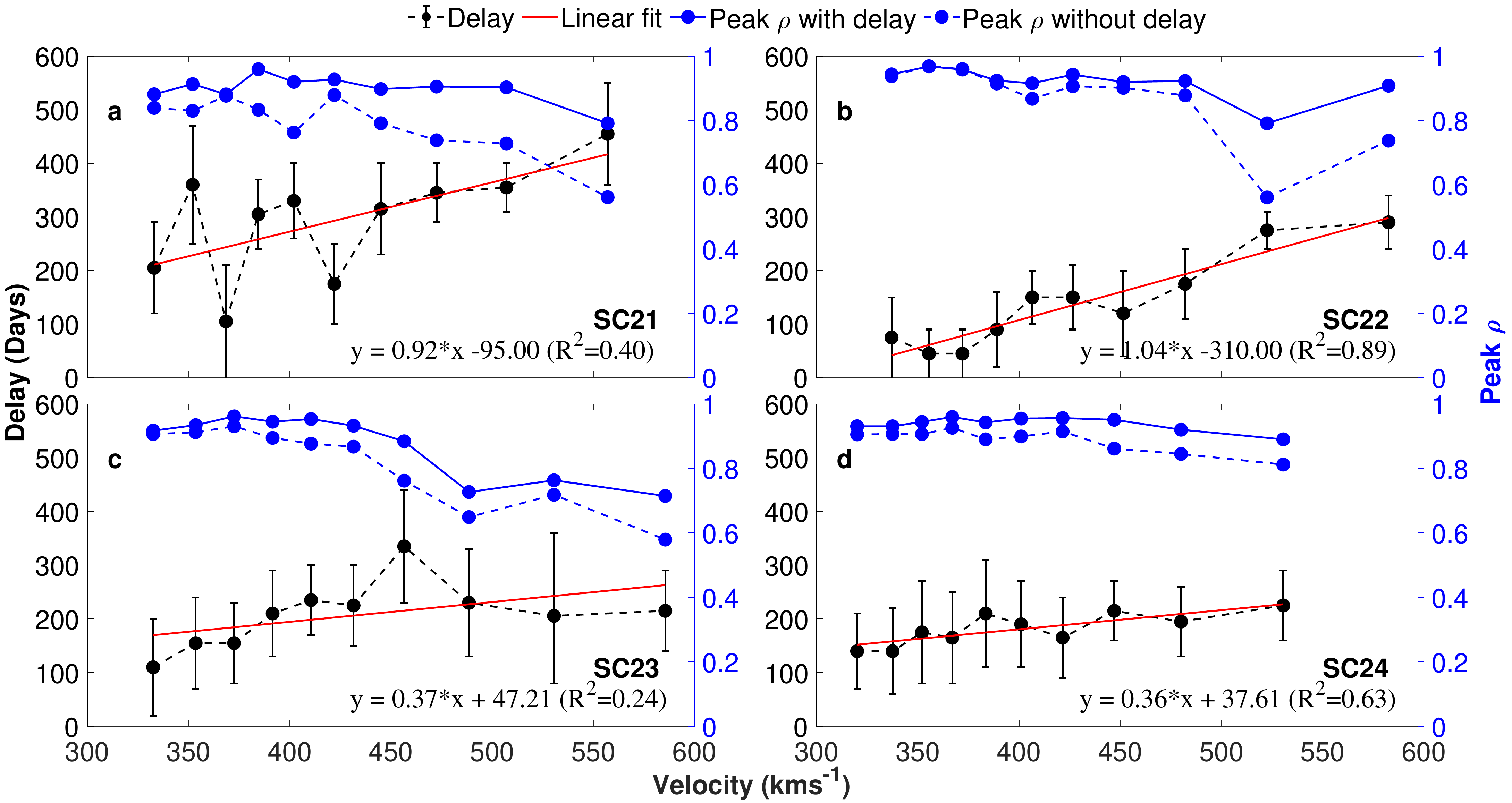}
\caption{\textbf{Delayed correlation between $A_{He}$ ($\leq$ 10\%) and SSN in four solar cycles.} Variation of delay (filled black dots joined by dashed line) between SSN and $A_{He}$ for each velocity quantile in four solar cycles marked by a (21), b(22), c(23) and d(24). The peak $\rho$ ($A_{He},SSN$) values between SSN and $A_{He}$ with (blue solid dots joined by solid blue line) and without (blue solid dots joined by dashed blue line) delay for each velocity quantile are also plotted for each cycle. The standard deviations are calculated by varying the peak $\rho$ ($A_{He},SSN$) by 1\% and considering the associated delays. The red lines mark the linear fits (equations and $R^2$ values are shown in each figure) for the delays in each cycle. The peak $\rho$ ($A_{He},SSN$) value maximizes between 350-400 km/s in all the four cycles. However, slopes of the linear fits in cycle 23-24 and particularly in cycle 24 are significantly less compared to cycles 21-22. \label{fig:4}}

\end{figure*}

In Fig. \ref{fig:3}, we divide the solar wind velocity in each cycle into 12 quantiles and neglect the fastest and slowest quantiles to avoid possible measurement uncertainties and spreading out of the velocity range. Therefore, this work is primarily applicable for slow and intermediate solar wind velocities. For each velocity bin, both $A_{He}$ and sunspot numbers (SSN) are first averaged for 250 days and plotted for the four cycles separately. The legend in each cycle shows the mid-point of each quantile and the corresponding Spearman Rank Correlation co-efficient, $\rho$ ($A_{He},SSN$), between $A_{He}$ and SSN. Fig, \ref{fig:3} reveals that $\rho$ ($A_{He},SSN$) maximizes at 369 km/s, 372 km/s, 373 km/s in cycles 21, 22 and 23 respectively. However, in cycle 24, The $\rho$ ($A_{He},SSN$) value maximizes at 401 km/s and it does not change significantly at higher velocity unlike other cycles.. Therefore, $\rho$ maximizes at a higher velocity in cycle 24 compared to the previous three solar cycle.

We also note that there is an apparent phase offset (delay) between the $A_{He}$ and SSN but this offset starts decreasing in cycle 23 and eventually becomes negligible in cycle 24. The dependence of this phase offset with the solar wind velocity is taken up for further scrutiny for the four solar cycles.

\subsection{Delay between $A_{He}$ and SSN}

In order to derive the phase offsets between $A_{He}$ and SSN for four solar cycles, both the time series are first subjected to 1-day average followed by 13 months smoothing. This is done to remove the discreteness in the data and also to eliminate the influence of the orientation of the heliospheric current sheet with respect to the satellites. SSNs are shifted in steps of 10 days starting from 0 to 600 and the peak $\rho$ ($A_{He},SSN$) values are obtained for each velocity bin with the corresponding delay times. These results are plotted in Fig. \ref{fig:4} that shows $\rho$ ($A_{He},SSN$) without delay in dashed blue line and peak $\rho$ ($A_{He},SSN$) with delay in solid blue line for each velocity quantile. The delays at different velocity quantiles are marked in filled black circles joined by dashed line to aid the eye. Positive delay means changes in $A_{He}$ follow the changes in SSN. The standard deviations are calculated by considering 1\% variation of the peak $\rho$ ($A_{He},SSN$) and the associated delays. It can be observed that the $\rho$ ($A_{He},SSN$) maximizes between 350-400 km/s in all the four cycles. However, the slopes of the linear fits (the fit equations are mentioned in the figure \ref{fig:4}) for delay are significantly reduced in cycle 23-24. We note that the delay for even the highest velocity quantile does not exceed 200 days. Not only that, the differences in delay between the lowest and highest quantiles are not significantly different. We varied $A_{He}$ from 4\% to 10\% and linear fit parameters are obtained in each case. This is tabulated and provided as a supplementary material. It is found that the slopes remain consistently higher (close to 1 or more) in cycle 21 and 22 and consistently lower (highest value of slope is 0.64 at $A_{He}$ = 5\%) in cycle 24 for all values of $A_{He}$.

\section{Discussion and Conclusions}

The helium abundance in the photosphere is taken to be nearly 8.5\% \citep{grevesse1998standard} but it generally remains about 4 – 5\% \citep{Laming2001Feldman} in the solar corona. This suggests that the helium abundance is depressed by processes that occur in the chromosphere, transition region and in the corona. Helium being heavier than hydrogen, it undergoes enhanced gravitational settling \citep{Hirshberg1973}. In addition, FIP effect \citep{Laming2015} can change the helium abundance. When the downward or upward propagating Alfv\'en waves encounter the chromosphere, these exert upward ponderomotive forces on the ions that raise the ions up in altitude \citep{Laming2012, Rakowski2012}. Subsequently, if the magnetic loops open up due to magnetic reconnection, the low FIP ions are released into the corona. While this effect enhances the abundance of lower FIP elements in the corona, it also conversely implies that it depletes the higher FIP elements. Therefore, FIP effect suggests that if there are significant changes in the closed loop coronal magnetic field configuration from one cycle to other, it will have impact on the helium abundance as well as delay.
  
This investigation shows that the maximum correlation coefficient between $A_{e}$ and SSN is obtained at a different velocity bin in cycle 24 as compared to the previous three cycles. Given this, it is also important to evaluate the sources of the slow/intermediate solar winds and how these affect the helium abundance. It was proposed earlier that the slow solar wind originates from two primary sources, viz., streamer belt and active region \citep{Kasper2007}. While active regions have stronger magnetic fields compared to streamer belt, the latter has longer magnetic loops that facilitates enhanced processing of the solar wind helium. This leads to the delay in $A_{He}$ with respect to the sunspot numbers. Interestingly, this delay is solar wind velocity dependent and, in general, more for higher velocities \citep{Alterman2019}. However, as shown by our analysis starting from cycle 23, this processing seems to have become less sensitive to the solar wind velocity. 
At this juncture, it may be noted that the time lag of the order of 100s of days suggests that He fractionation is a slow process. Whether it indicates the dominant role of gravitational settling in the He fractionation, is a question that requires further attention. Interestingly, as discussed by \cite{Laming2019}, extreme He depletions reported by earlier works \citep{Kasper2012, Kepko2016} are unlikely to be reproduced by ponderomotive force and thus gravitational settling appears to be the most dominant mechanism. Therefore, the reduced time lag across the velocity bins and systematic shift of $A_{He}$ frequency to lower $A_{He}$ values in cycle 24 may be suggestive of systematic changes in ponderomotive force as well as the gravitational settling in closed magnetic loops. It is to be noted that length of the magnetic loop is also important for He processing. Although these are speculative scenarios, there are indications that systematic changes did occur in the Sun starting from cycle 23. One such example is the changes in the occurrence and characteristics of pseudostreamers and dipole streamers. Pseudostreamers are known to be additional sources of slow solar wind \citep{Crooker2012}.  Interestingly, occurrence of pseudostreamers were found to be significantly higher during the minimum of the solar cycle 23 unlike cycles 21 and 22 when these structures almost disappeared \citep{Owens2014}.  On the other hand, the dipole streamers were shown \citep{Owens2014} to be less tightly confined to the solar equator at the end of cycle 23 compared to the previous two cycles. This difference in the pseudostreamers and dipole streamers at the end of solar cycle 23 indicate towards the possible change in the processing of the slow solar wind at the end of cycle 23.

 Some of the earlier works \citep{Levine1977, Wang1990} suggest that the origin of slow solar wind is the boundary of the coronal holes. As the magnetic flux tubes from the boundary of the coronal hole are expanded, the bulk speed of the slow solar wind decreases. This reduces the proton flux beyond 2.5 $R_\odot$ and as a result, the alpha particle enrichment in the solar wind decreases through reduced momentum transfer via Coulomb collisions \citep{Burgi1992}. This scenario is particularly valid for solar minimum when the magnetic flux tubes undergo strong over-expansion. We suggest that large scale magnetic flux tube topologies are quite different (in statistical sense) in solar cycle 24 (starting from 23) that affected the processing as well as the velocity dependence of the helium enrichment process in the solar wind. This proposition gets indirect support from the fact that the polar field went through unusual reversal in SC24 \citep{Janardhan2018}.  

The higher $A_{He}$ events is, in general, thought to be associated with transient coronal disturbances like solar flare, eruptive prominences etc \citep{Borrini1982}. Many of these processes are believed to throw helium-rich plasma from the lower solar corona to the interplanetary medium causing helium abundance enhancements. The frequency of occurrence of such transient disturbances is approximately in phase with the solar cycle \citep{Borrini1982}. Further, it is known that the number of X, M and C-class flares dropped drastically in cycle 23 compared to the earlier cycles  \citep{Hudson2014}. The number of such flares is even less in cycle 24. This may be one of the reasons for the lower frequency of higher $A_{He}$ events in cycle 24 compared to the previous three cycles. 
As stated earlier, solar cycle 24 turns out to be the weakest in the last century and the Sun was unusually quiet since the deep minimum of cycle 23.  Many authors highlighted the declining solar activity from the cycle 23 onwards based on observations \citep{Janardhan2011} or dynamo models  \citep{Choudhuri2007}. Observations during Ulysses first (cycle 22) and third orbit (cycle 23) revealed that solar wind in the minimum of cycle 23 was weaker characterized by significantly less dense, cooler wind with less mass and momentum flux compared to the first orbit during the solar minimum in cycle 22 \citep{McComas2008}. The peculiarities of solar cycle 24 have also been highlighted by earlier workers \citep{Antia2013}. Therefore, we believe that the large scale coronal magnetic field topology and dynamics of the Sun went through significant changes in cycle 24 that affected the way helium was processed in the solar wind. Recent helium abundance measurements \citep{Moses2020} in the solar corona provides credence to this proposition. Further modelling investigations are needed to explain the changes in the helium abundance in cycle 24 and to predict the possible variation in the cycle 25.

\section*{Acknowledgements}

We would like to thank the managing team of OMNI dataset (\url{https://cdaweb.gsfc.nasa.gov/team.html}). We thank and acknowledge the significant efforts put forward by the PIs of the all satellites, data of which are used to generate this integrated, cross-calibrated OMNI dataset. This work is supported by the Department of Space, Government of India.

\section*{Data Availability}
The data can be obtained from \url{https://cdaweb.gsfc.nasa.gov/index.html/}.



\bibliographystyle{mnras}
\bibliography{cite_MN-21-0020-L.R1} 

\begin{thebibliography}{}
\makeatletter
\relax
\def\mn@urlcharsother{\let\do\@makeother \do\$\do\&\do\#\do\^\do\_\do\%\do\~}
\def\mn@doi{\begingroup\mn@urlcharsother \@ifnextchar [ {\mn@doi@}
  {\mn@doi@[]}}
\def\mn@doi@[#1]#2{\def\@tempa{#1}\ifx\@tempa\@empty \href
  {http://dx.doi.org/#2} {doi:#2}\else \href {http://dx.doi.org/#2} {#1}\fi
  \endgroup}
\def\mn@eprint#1#2{\mn@eprint@#1:#2::\@nil}
\def\mn@eprint@arXiv#1{\href {http://arxiv.org/abs/#1} {{\tt arXiv:#1}}}
\def\mn@eprint@dblp#1{\href {http://dblp.uni-trier.de/rec/bibtex/#1.xml}
  {dblp:#1}}
\def\mn@eprint@#1:#2:#3:#4\@nil{\def\@tempa {#1}\def\@tempb {#2}\def\@tempc
  {#3}\ifx \@tempc \@empty \let \@tempc \@tempb \let \@tempb \@tempa \fi \ifx
  \@tempb \@empty \def\@tempb {arXiv}\fi \@ifundefined
  {mn@eprint@\@tempb}{\@tempb:\@tempc}{\expandafter \expandafter \csname
  mn@eprint@\@tempb\endcsname \expandafter{\@tempc}}}

\bibitem[\protect\citeauthoryear{Aellig, Lazarus  \& Steinberg}{Aellig
  et~al.}{2001}]{Aellig2001}
Aellig M.~R.,  Lazarus A.~J.,   Steinberg J.~T.,  2001, \mn@doi [AIP Conference
  Proceedings] {10.1063/1.1433984}, 598, 89

\bibitem[\protect\citeauthoryear{Alterman \& Kasper}{Alterman \&
  Kasper}{2019}]{Alterman2019}
Alterman B.~L.,  Kasper J.~C.,  2019, \mn@doi [The Astrophysical Journal]
  {10.3847/2041-8213/ab2391}, 879, L6

\bibitem[\protect\citeauthoryear{Antia \& Basu}{Antia \&
  Basu}{2013}]{Antia2013}
Antia H.~M.,  Basu S.,  2013, \mn@doi [Journal of Physics: Conference Series]
  {10.1088/1742-6596/440/1/012018}, 440, 012018

\bibitem[\protect\citeauthoryear{Borrini, Gosling, Bame  \& Feldman}{Borrini
  et~al.}{1982}]{Borrini1982}
Borrini G.,  Gosling J.~T.,  Bame S.~J.,   Feldman W.~C.,  1982, \mn@doi
  [Journal of Geophysical Research: Space Physics] {10.1029/JA087iA09p07370},
  87, 7370

\bibitem[\protect\citeauthoryear{Bürgi}{Bürgi}{1992}]{Burgi1992}
Bürgi A.,  1992, \mn@doi [Journal of Geophysical Research: Space Physics]
  {10.1029/91JA02833}, 97, 3137

\bibitem[\protect\citeauthoryear{Choudhuri, Chatterjee  \& Jiang}{Choudhuri
  et~al.}{2007}]{Choudhuri2007}
Choudhuri A.~R.,  Chatterjee P.,   Jiang J.,  2007, \mn@doi [Phys. Rev. Lett.]
  {10.1103/PhysRevLett.98.131103}, 98, 131103

\bibitem[\protect\citeauthoryear{Crooker, Antiochos, Zhao  \&
  Neugebauer}{Crooker et~al.}{2012}]{Crooker2012}
Crooker N.~U.,  Antiochos S.~K.,  Zhao X.,   Neugebauer M.,  2012, \mn@doi
  [Journal of Geophysical Research: Space Physics]
  {https://doi.org/10.1029/2011JA017236}, 117

\bibitem[\protect\citeauthoryear{Feldman, Asbridge, Bame  \& Gosling}{Feldman
  et~al.}{1978}]{feldman1978long}
Feldman W.,  Asbridge J.,  Bame S.,   Gosling J.,  1978, \mn@doi [Journal of
  Geophysical Research: Space Physics] {10.1007/BF00148286}, \href
  {https://doi.org/10.1007/BF00148286} {83, 2177}

\bibitem[\protect\citeauthoryear{Grevesse \& Sauval}{Grevesse \&
  Sauval}{1998}]{grevesse1998standard}
Grevesse N.,  Sauval A.,  1998, \mn@doi [Space Science Reviews]
  {10.1023/A:1005161325181}, 85, 161

\bibitem[\protect\citeauthoryear{Hathaway}{Hathaway}{2015}]{hathaway2015solar}
Hathaway D.~H.,  2015, Living reviews in solar physics, 12, 4

\bibitem[\protect\citeauthoryear{{Hirshberg}}{{Hirshberg}}{1973}]{Hirshberg1973}
{Hirshberg} J.,  1973, \mn@doi [Reviews of Geophysics and Space Physics]
  {10.1029/RG011i001p00115}, 11, 115

\bibitem[\protect\citeauthoryear{Hudson, Fletcher  \& McTiernan}{Hudson
  et~al.}{2014}]{Hudson2014}
Hudson H.,  Fletcher L.,   McTiernan J.,  2014, \mn@doi [Solar Physics]
  {10.1007/s11207-013-0384-7}, 289, 1341

\bibitem[\protect\citeauthoryear{Janardhan, Bisoi, Ananthakrishnan, Tokumaru
  \& Fujiki}{Janardhan et~al.}{2011}]{Janardhan2011}
Janardhan P.,  Bisoi S.~K.,  Ananthakrishnan S.,  Tokumaru M.,   Fujiki K.,
  2011, \mn@doi [Geophysical Research Letters] {10.1029/2011GL049227}, 38

\bibitem[\protect\citeauthoryear{Janardhan, Fujiki, Ingale, Bisoi  \&
  Rout}{Janardhan et~al.}{2018}]{Janardhan2018}
Janardhan P.,  Fujiki K.,  Ingale M.,  Bisoi S.~K.,   Rout D.,  2018, \mn@doi
  [Astronomy & Astrophysics] {10.1051/0004-6361/201832981}, 618, A148

\bibitem[\protect\citeauthoryear{{Kasper}, {Stevens}, {Lazarus}, {Steinberg}
  \& {Ogilvie}}{{Kasper} et~al.}{2007}]{Kasper2007}
{Kasper} J.~C.,  {Stevens} M.~L.,  {Lazarus} A.~J.,  {Steinberg} J.~T.,
  {Ogilvie} K.~W.,  2007, \mn@doi [The Astrophysical Journal] {10.1086/510842},
  660, 901

\bibitem[\protect\citeauthoryear{{Kasper}, {Stevens}, {Korreck}, {Maruca},
  {Kiefer}, {Schwadron}  \& {Lepri}}{{Kasper} et~al.}{2012}]{Kasper2012}
{Kasper} J.~C.,  {Stevens} M.~L.,  {Korreck} K.~E.,  {Maruca} B.~A.,  {Kiefer}
  K.~K.,  {Schwadron} N.~A.,   {Lepri} S.~T.,  2012, \mn@doi [The Astrophysical
  Journal] {10.1088/0004-637X/745/2/162}, 745, 162

\bibitem[\protect\citeauthoryear{Kepko, Viall, Antiochos, Lepri, Kasper  \&
  Weberg}{Kepko et~al.}{2016}]{Kepko2016}
Kepko L.,  Viall N.~M.,  Antiochos S.~K.,  Lepri S.~T.,  Kasper J.~C.,   Weberg
  M.,  2016, \mn@doi [Geophysical Research Letters]
  {https://doi.org/10.1002/2016GL068607}, 43, 4089

\bibitem[\protect\citeauthoryear{King \& Papitashvili}{King \&
  Papitashvili}{2005}]{King2005}
King J.~H.,  Papitashvili N.~E.,  2005, \mn@doi [Journal of Geophysical
  Research: Space Physics] {10.1029/2004JA010649}, 110

\bibitem[\protect\citeauthoryear{{Laming}}{{Laming}}{2012}]{Laming2012}
{Laming} J.~M.,  2012, \mn@doi [The Astrophysical Journal]
  {10.1088/0004-637X/744/2/115}, 744, 115

\bibitem[\protect\citeauthoryear{{Laming}}{{Laming}}{2015}]{Laming2015}
{Laming} J.~M.,  2015, \mn@doi [Living Reviews in Solar Physics]
  {10.1007/lrsp-2015-2}, 12, 2

\bibitem[\protect\citeauthoryear{Laming \& Feldman}{Laming \&
  Feldman}{2001}]{Laming2001Feldman}
Laming J.~M.,  Feldman U.,  2001, \mn@doi [The Astrophysical Journal]
  {10.1086/318238}, 546, 552

\bibitem[\protect\citeauthoryear{{Laming} et~al.,}{{Laming}
  et~al.}{2019}]{Laming2019}
{Laming} J.~M.,  et~al., 2019, \mn@doi [\apj] {10.3847/1538-4357/ab23f1}, \href
  {https://ui.adsabs.harvard.edu/abs/2019ApJ...879..124L} {879, 124}

\bibitem[\protect\citeauthoryear{{Levine}, {Altschuler}  \& {Harvey}}{{Levine}
  et~al.}{1977}]{Levine1977}
{Levine} R.~H.,  {Altschuler} M.~D.,   {Harvey} J.~W.,  1977, \mn@doi [Journal
  of Geophysical Research] {10.1029/JA082i007p01061}, 82, 1061

\bibitem[\protect\citeauthoryear{McComas, Ebert, Elliott, Goldstein, Gosling,
  Schwadron  \& Skoug}{McComas et~al.}{2008}]{McComas2008}
McComas D.~J.,  Ebert R.~W.,  Elliott H.~A.,  Goldstein B.~E.,  Gosling J.~T.,
  Schwadron N.~A.,   Skoug R.~M.,  2008, \mn@doi [Geophysical Research Letters]
  {10.1029/2008GL034896}, 35

\bibitem[\protect\citeauthoryear{Moses et~al.,}{Moses et~al.}{2020}]{Moses2020}
Moses J.,  et~al., 2020, Nature Astronomy

\bibitem[\protect\citeauthoryear{Ogilvie \& Hirshberg}{Ogilvie \&
  Hirshberg}{1974}]{Ogilvie1974}
Ogilvie K.~W.,  Hirshberg J.,  1974, \mn@doi [Journal of Geophysical Research
  Space Physics] {10.1029/JA079i031p04595}, 79, 4595

\bibitem[\protect\citeauthoryear{Owens, Crooker  \& Lockwood}{Owens
  et~al.}{2014}]{Owens2014}
Owens M.~J.,  Crooker N.~U.,   Lockwood M.,  2014, \mn@doi [Journal of
  Geophysical Research: Space Physics] {https://doi.org/10.1002/2013JA019412},
  119, 36

\bibitem[\protect\citeauthoryear{Rakowski \& Laming}{Rakowski \&
  Laming}{2012}]{Rakowski2012}
Rakowski C.~E.,  Laming J.~M.,  2012, \mn@doi [The Astrophysical Journal]
  {10.1088/0004-637x/754/1/65}, 754, 65

\bibitem[\protect\citeauthoryear{{Wang} \& {Sheeley}}{{Wang} \&
  {Sheeley}}{1990}]{Wang1990}
{Wang} Y.~M.,  {Sheeley} N.~R. J.,  1990, \mn@doi [The Astrophysical Journal]
  {10.1086/168805}, 355, 726

\makeatother
\end{thebibliography}



\label{lastpage}

\end{document}


\begin{table}[ht]
\caption{\textbf{Linear fit parameters between the delay and velocity for different $A_{He}$ limits:} The limit of $A_{He}$ is varied from 4\% to 10\% and the corresponding delays are estimated. The $A_{He}$ limit of 10\%  means all the $A_{He}$ data upto $A_{He}$ = 10\% are considered The linear fit parameters  are stated in the table. The fit parameters include  slope (m), intercept (c) and  the goodness of fit coefficient ($R^2$). The rows corresponding to the slope in each cycle are marked in bold fonts. It is found that the slopes remain consistently higher (close to 1 or more) in cycle 21 and 22 and consistently lower (highest value of slope is 0.64 at $A_{He}$ = 5\%) in cycle 24 for all values of $A_{He}$.\label{tab:1}}
\begin{center}
\begin{tabular}{|c|c|c|c|c|c|c|c|c|}
 \hline
SC  & Modelled & \multicolumn{7}{c|}{$A_{He}$ limit} \\ 
 \cline{3-9}
  & parameters  & 4\% & 5\% & 6\% & 7\% & 8\% & 9\% & 10\% \\
  \hline
  & \textbf{m} & \textbf{1.85}& \textbf{1.12}&	\textbf{1.34}&	\textbf{1.58}&	\textbf{1.35} & \textbf{1.11} & \textbf{0.92}\\ 
\cline{2-9} 
21 & c &-508.24 & -56.89	&	-157.70&-285.37	&-262.95 & -143.45&94.93\\ 
\cline{2-9}
 &  $R^2$ &  0.39& 0.22&	0.42&	0.54& 0.61 & 0.61&0.40\\ 
 \hline
   & \textbf{m} &\textbf{2.06}&	\textbf{2.22}&	\textbf{2.04}&	\textbf{1.69}&	\textbf{1.27} &\textbf{1.08} &\textbf{1.04}\\ 
\cline{2-9} 
 22 & c & -671.54&	-756.51 &	-688.91&	-543.61&	-389.24& -310.99&-309.99\\ 
\cline{2-9}
 & $R^2$ &  0.75& 0.0.84& 0.82& 0.64& 0.86 &0.85 &0.89\\ 
 \hline 
 & \textbf{m} & \textbf{0.99}&	\textbf{1.90}&	\textbf{1.66}&	\textbf{1.95}&	\textbf{0.60} &\textbf{0.30} &\textbf{0.30}\\ 
\cline{2-9} 
23 & c &	-153.54&	-498.31&	-419.18&	-531.84 &-28.42 &78.65 & 47.21\\ 
\cline{2-9}
 & $R^2$ & 0.56& 0.97& 	0.75&	0.94& 0.35 &0.16 &0.24\\ 
 \hline 
 & \textbf{m} & \textbf{0.35}&	\textbf{0.64}&	\textbf{0.62}&	\textbf{0.55}&	\textbf{0.43} &\textbf{0.32} &\textbf{0.36}\\ 
\cline{2-9} 
24  & c &55.19&	-54.05&	-50.91&	-22.36&	16.14 &56.87 &37.61\\ 
\cline{2-9} 
  & $R^2$ & 0.21 & 0.81 & 0.69& 0.50 &0.63 & 0.41  &0.63\\ 
 \hline 
\end{tabular}

\end{center}
\end{table}